\documentclass[aps,pra,superscriptaddress]{revtex4}
\usepackage{graphics}
\usepackage{amsmath}
\usepackage{graphicx}
\usepackage{amsfonts}
\usepackage{amssymb}

\begin{document}

\title{On the security and degradability of Gaussian channels}
\author{Stefano Pirandola}
\affiliation{Research Laboratory of Electronics, MIT, Cambridge MA 02139, USA}
\author{Samuel L. Braunstein}
\affiliation{Department of Computer Science, University of York, York YO10 5DD, UK}
\author{Seth Lloyd}
\affiliation{Research Laboratory of Electronics, MIT, Cambridge MA 02139, USA}
\affiliation{Department of Mechanical Engineering, MIT, Cambridge MA 02139, USA}
\date{\today }

\begin{abstract}
We consider the notion of canonical attacks, which are the cryptographic
analog of the canonical forms of a one-mode Gaussian channel. Using this
notion, we explore the connections between the degradability properties of
the channel and its security for quantum key distribution. Finally, we also
show some relations between canonical attacks and optimal Gaussian cloners.
\end{abstract}

\maketitle

%

\section{Introduction}

Today, quantum cryptography is one of the most promising areas in quantum
information science. This is particularly true in the framework of
continuous variable (CV) systems \cite{cvBook}, which are quantum systems
characterized by infinite-dimensional Hilbert spaces. The increasing
interest in CV quantum cryptography is mainly due to the practical
advantages of quantum key distribution (QKD) using Gaussian states \cite%
{CVQKD,Homo2,Hetero2,GaussianATT,reverseSKC}. Furthermore, this Gaussian
QKD\ has been also extended to multiple quantum communications \cite{twoWAY}
and the non-trivial possibility of a quantum direct communication has been
also explored \cite{QDC}. Very recently, a new insight in the theory of
quantum channels has been provided by the canonical classification of the
one-mode Gaussian channels \cite{HOL} (see also Refs.~\cite{Jens,VCH}\ and
the \textit{compact version} of this classification in Ref.~\cite%
{GaussianATT}). These channels have been proven to be unitarily equivalent
to canonical forms of six different classes \cite{HOL}, whose degradability
properties have been also studied \cite{VCH}. Here, we exploit these
concepts in the scenario of quantum cryptography. In particular, we consider
the notion of canonical attacks as the cryptographic analog of the canonical
forms. By adopting the individual version of these canonical attacks, we
explore the connections between the degradability properties of the channel
and its security for QKD. Then, we also show when (and in what sense) these
attacks can be considered equivalent to individual attacks using optimal
Gaussian cloners.

\section{Quantum communication scenario}

The simplest continuous variable system is a single bosonic mode, i.e., a
quantum system described by a pair of quadrature operators $\mathbf{\hat{x}}%
^{T}:=(\hat{q},\hat{p})$ with $[\hat{q},\hat{p}]=2i$. In particular, a
single-mode bosonic state $\rho $ with Gaussian statistics is called \emph{%
Gaussian state} and it is completely characterized by a $2\times 2$
covariance matrix $\mathbf{V}$ plus a displacement vector $\mathbf{\bar{x}}%
\in \mathbb{R}^{2}$. Then, a one-mode Gaussian channel is a completely
positive trace-preserving (CPT) map $\mathcal{G}(\mathbf{T},\mathbf{N},%
\mathbf{d})$ transforming an input Gaussian state $\rho _{a}(\mathbf{V}_{a},%
\mathbf{\bar{x}}_{a})$ of a sender (Alice) into an output Gaussian state $%
\rho _{b}(\mathbf{V}_{b},\mathbf{\bar{x}}_{b})$ of a receiver (Bob) via the
relations $\mathbf{V}_{b}=\mathbf{TV}_{a}\mathbf{T}^{T}+\mathbf{N}$ and $%
\mathbf{\bar{x}}_{b}=\mathbf{T\bar{x}}_{a}\mathbf{+d}$. Here, $\mathbf{d}\in
\mathbb{R}^{2}$ and $\mathbf{T},\mathbf{N}$ are $2\times 2$ real matrices,
with $\mathbf{N}^{T}=\mathbf{N}>0$ and $\det \mathbf{N}\geq \left( \det
\mathbf{T}-1\right) ^{2}$. Up to unitaries on the input and the output,
every one-mode Gaussian channel is equivalent to a map $\mathcal{C}$, called
the \emph{canonical form}, which is a Gaussian channel with $\mathbf{d}=%
\mathbf{0}$ and $\mathbf{T}_{c},\mathbf{N}_{c}$ diagonal \cite{HOL}.
According to Ref.~\cite{GaussianATT}, the explicit expressions of $\mathbf{T}%
_{c}$ and $\mathbf{N}_{c}$ depend on three symplectic invariants of the
channel: the generalized \emph{transmission} $\tau :=\det \mathbf{T}$
(ranging from $-\infty $ to $+\infty $), the \emph{rank} $r:=[$rk$(\mathbf{T}%
)$rk$(\mathbf{N})]/2$ (with possible values $r=0,1,2$) and the \emph{%
temperature }$\bar{n}$ (which is a positive number related to $\det \mathbf{N%
}$ \cite{GaussianATT}). These three invariants $\{\tau ,r,\bar{n}\}$
completely characterize the two matrices $\mathbf{T}_{c},\mathbf{N}_{c}$
and, therefore, the corresponding canonical form $\mathcal{C}=\mathcal{C}%
(\tau ,r,\bar{n})$. In particular, the first two invariants $\{\tau ,r\}$
determine the class of the form \cite{GaussianATT,HOL}. The full
classification is explicitly shown in the following table%
\begin{equation*}
\begin{tabular}{c|c||c||c||c|c}
$\tau $ & $~r~$ & Class & ~~~Form~~ & $\mathbf{T}_{c}$ & $\mathbf{N}_{c}$ \\
\hline
$0$ & $0$ & $A_{1}$ & $\mathcal{C}(0,0,\bar{n})$ & $\mathbf{0}$ & $(2\bar{n}%
+1)\mathbf{I}$ \\
$0$ & $1$ & $A_{2}$ & $\mathcal{C}(0,1,\bar{n})$ & $\frac{\mathbf{I}+\mathbf{%
Z}}{2}$ & $(2\bar{n}+1)\mathbf{I}$ \\
$1$ & $1$ & $B_{1}$ & $\mathcal{C}(1,1,0)$ & $\mathbf{I}$ & $\frac{\mathbf{I}%
-\mathbf{Z}}{2}$ \\
$1$ & $2$ & $B_{2}$ & $\mathcal{C}(1,2,\bar{n})$ & $\mathbf{I}$ & $\bar{n}%
\mathbf{I}$ \\
$1$ & $0$ & $B_{2}(Id)$ & $\mathcal{C}(1,0,0)$ & $\mathbf{I}$ & $\mathbf{0}$
\\
$(0,1)$ & $2$ & $C(Att)$ & $\mathcal{C}(\tau ,2,\bar{n})$ & $\sqrt{\tau }%
\mathbf{I}$ & $(1-\tau )(2\bar{n}+1)\mathbf{I}$ \\
$>1$ & $2$ & $C(Amp)$ & $\mathcal{C}(\tau ,2,\bar{n})$ & $\sqrt{\tau }%
\mathbf{I}$ & $(\tau -1)(2\bar{n}+1)\mathbf{I}$ \\
$<0$ & $2$ & $D$ & $\mathcal{C}(\tau ,2,\bar{n})$ & $\sqrt{-\tau }\mathbf{Z}$
& $(1-\tau )(2\bar{n}+1)\mathbf{I}$%
\end{tabular}%
\end{equation*}%
In this table, the values of $\{\tau ,r\}$ in the first two columns specify
a particular class $A_{1},A_{2},B_{1},B_{2},C$ and $D$ \cite{Table}. Within
each class, the possible canonical forms are expressed in the third column,
where also the third invariant $\bar{n}$ must be considered. The
corresponding expressions of $\mathbf{T}_{c},\mathbf{N}_{c}$ are shown in
the last two columns, where $\mathbf{Z}:=\mathrm{diag}(1,-1)$, $\mathbf{I}:=%
\mathrm{diag}(1,1)$ and $\mathbf{0}$ is the zero matrix.

\begin{figure}[tbph]
\vspace{-2.0cm}
\par
\begin{center}
\includegraphics[width=0.91\textwidth] {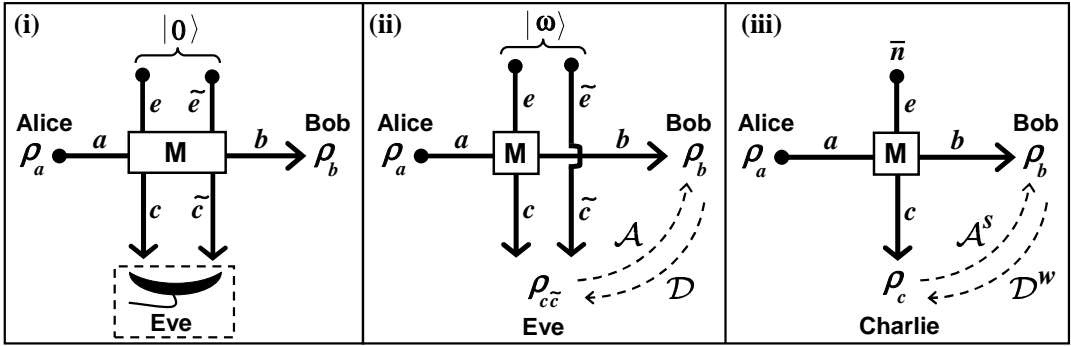}
\end{center}
\par
\vspace{-3.0cm}
\caption{Inset (i): Stinespring dilation of a canonical form. It describes a
canonical attack by including the optimal coherent detection of all the
outputs $\{c,\tilde{c}\}$ collected in all the uses of the channel. Both the
dilation and the attack are unique up to isometries acting on the
environmental modes $\{c,\tilde{c}\}$. Inset (ii): Stinespring dilation of a
canonical form of every class but $B_{2}$; the form\ is antidegradable
(degradable) if there exists a map $\mathcal{A}$ ($\mathcal{D}$) such that $%
\protect\rho _{b}=\mathcal{A}(\protect\rho _{c\tilde{c}})$\ [$\protect\rho %
_{c\tilde{c}}=\mathcal{D}(\protect\rho _{b})$]. Inset (iii): Physical
representation of a canonical form of every class but $B_{2}$; the form\ is
strongly antidegradable (weakly degradable) if there exists a map $\mathcal{A%
}^{s}$ ($\mathcal{D}^{w}$) such that $\protect\rho _{b}=\mathcal{A}^{s}(%
\protect\rho _{c})$\ [$\protect\rho _{c}=\mathcal{D}^{w}(\protect\rho _{b})$%
].}
\label{CharlieEvepic}
\end{figure}

By adopting a Stinespring dilation of the quantum channel, we can describe a
canonical form $\mathcal{C}(\tau ,r,\bar{n})$ via a symplectic
transformation $\mathbf{M}_{ae\tilde{e}}$ mixing the input signal mode $%
\{a\} $\ with two vacuum environmental modes $\{e,\tilde{e}\}$ and yielding
the output modes $\{b\}$ for Bob and $\{c,\tilde{c}\}$ for the environment
[see Fig.~\ref{CharlieEvepic}(i)]. Such a dilation is known to be unique up
to partial isometries. For class $B_{2}$, also known as an \emph{%
additive-noise channel}, the Stinespring dilation corresponds to an optimal
Gaussian cloner (OGC) which clones asymmetrically in the clones but
symmetrically in the quadratures \cite{Cerf}. Such a machine transforms the
input mode $\{a\}$ and the two vacuum modes $\{e,\tilde{e}\}$ into a pair of
clone modes $\{b,c\}$ and an anticlone mode $\{\tilde{c}\}$. In particular
the reduced state of the output clone $k=b,c$ is given by the modulated
state
\begin{equation}
\rho _{k}=\int d^{2}\gamma ~G_{\chi _{k}}(\gamma )\hat{D}(\gamma )\rho _{a}%
\hat{D}^{\dag }(\gamma )~,~G_{\chi _{k}}(\gamma ):=\frac{1}{\pi \chi _{k}}%
\exp \left( -\frac{\left\vert \gamma \right\vert ^{2}}{\chi _{k}}\right) ~,
\end{equation}%
where $\hat{D}(\gamma )$ is the displacement operator and $\chi _{k}$ is the
cloning noise variance satisfying $\chi _{b}\chi _{c}=1$ with $\chi _{b}=%
\bar{n}$.

It is important to notice that class $B_{2}$ is the unique class where the
Stinespring dilation cannot be simplified to a single-mode description \cite%
{HOL}. For all the other classes, in fact, we can consider a dilation where
the symplectic transformation $\mathbf{M}_{ae\tilde{e}}$ can be decomposed
as $\mathbf{M}_{ae}\oplus \mathbf{I}_{\tilde{e}}$, involving the signal mode
$\{a\}$\ and \emph{only one} mode $\{e\}$\ of the two-mode environment $\{e,%
\tilde{e}\}$ [see Fig.~\ref{CharlieEvepic}(ii)]. This is possible if the
environment is prepared in a two-mode squeezed vacuum (TMSV) state $%
\left\vert \omega \right\rangle _{e\tilde{e}}$, i.e., in a pure Gaussian
state with correlation matrix%
\begin{equation}
\mathbf{V}_{e\tilde{e}}(\omega )=\left(
\begin{array}{cc}
\omega \mathbf{I} & \sqrt{\omega ^{2}-1}\mathbf{Z} \\
\sqrt{\omega ^{2}-1}\mathbf{Z} & \omega \mathbf{I}%
\end{array}%
\right) ~,~\omega :=2\bar{n}+1\geq 1~.  \label{Input_EPR}
\end{equation}%
Such a dilation $\{\mathbf{M}_{ae}\oplus \mathbf{I}_{\tilde{e}},\left\vert
\omega \right\rangle \}$ is the purification of a single-mode \textit{%
physical representation} \cite{PRnote} $\{\mathbf{M}_{ae},\rho (\bar{n})\}$,
where $\mathbf{M}_{ae}$ mixes the signal mode $\{a\}$ with a single
environmental mode $\{e\}$, prepared in a thermal state $\rho _{e}(\bar{n})$
with $\bar{n}$ average photons \cite{VCH}. This physical representation can
also be seen as a quantum broadcast channel where the symplectic $\mathbf{M}%
_{ae}$ relates the output quadratures $\mathbf{\hat{x}}_{out}^{T}=(\hat{q}%
_{b},\hat{p}_{b},\hat{q}_{c},\hat{p}_{c})$ of two receivers (Bob and
Charlie) to the input quadratures $\mathbf{\hat{x}}_{in}^{T}=(\hat{q}_{a},%
\hat{p}_{a},\hat{q}_{e},\hat{p}_{e})$ of a sender (Alice) and a thermal
environment [see Fig.~\ref{CharlieEvepic}(iii)].

An important property of all the canonical forms (except $B_{2}$ \cite{B2})
is their \emph{degradability }or \emph{antidegradability} \cite{VCH,DeveShor}%
. A canonical form is called \textit{strongly antidegradable} if it has a
single-mode physical representation $\{\mathbf{M}_{ae},\rho (\bar{n})\}$
where Charlie can reconstruct Bob's state $\rho _{b}$ via some CPT map $%
\mathcal{A}^{s}$, i.e., $\rho _{b}=\mathcal{A}^{s}(\rho _{c})$. Notice that
the strong antidegradability $\{a\}\rightarrow \{c\}\rightarrow \{b\}$ is a
sufficient but not necessary condition for the (standard) antidegradability $%
\{a\}\rightarrow \{c,\tilde{c}\}\rightarrow \{b\}$, where Bob's state is
reconstructed by considering all the degrees of freedom of the environment
(Eve). In fact, the form is called \textit{antidegradable} if there exists a
CPT\ map $\mathcal{A}$ such that $\rho _{b}=\mathcal{A}(\rho _{c\tilde{c}})$%
.\ In the same way, one can consider the \textit{weak degradability }$%
\{a\}\rightarrow \{b\}\rightarrow \{c\}$ which is implied by the (standard)
degradability $\{a\}\rightarrow \{b\}\rightarrow \{c,\tilde{c}\}$. In fact,
degradability corresponds to the existence of a map $\mathcal{D}$ such that $%
\rho _{c\tilde{c}}=\mathcal{D}(\rho _{b})$, while weak degradability
corresponds to the existence of a map $\mathcal{D}^{w}$ such that $\rho _{c}=%
\mathcal{D}^{w}(\rho _{b})$ for some physical representation. Clearly the
weak/strong notions coincide with the standard ones when $\bar{n}=1$.

From the point of view of practical quantum cryptography, classes $C$ and $D$
are the most important ones. These classes are full-rank ($r=2$) and
represent the unique classes where the invariant $\tau $ can take a
continuum of values, except for the singular points $\tau =0$ and $\tau =1$.
Because of this continuity, we call the canonical forms $\mathcal{C}(\tau ,2,%
\bar{n})$ of classes $C$ and $D$ as \emph{regular}. For these forms, one can
consider a single-mode physical representation $\{\mathbf{M}_{ae},\rho (\bar{%
n})\}$ with symplectic matrix%
\begin{equation}
\mathbf{M}_{ae}(\tau )=\left(
\begin{array}{cccc}
\sqrt{\left\vert \tau \right\vert } & 0 & \sqrt{\left\vert 1-\tau
\right\vert } & 0 \\
0 & s(\tau )\sqrt{\left\vert \tau \right\vert } & 0 & s(1-\tau )\sqrt{%
\left\vert 1-\tau \right\vert } \\
s(\tau -1)\sqrt{\left\vert 1-\tau \right\vert } & 0 & s(\tau )\sqrt{%
\left\vert \tau \right\vert } & 0 \\
0 & -\sqrt{\left\vert 1-\tau \right\vert } & 0 & \sqrt{\left\vert \tau
\right\vert }%
\end{array}%
\right) ~,  \label{Symplectic_Interaction}
\end{equation}%
where $s(\cdots )$\ is the sign function. Notice that Eq.~(\ref%
{Symplectic_Interaction}) corresponds to a beam splitter for $0<\tau <1$ and
to an amplifier for $\tau >1$. A regular canonical form $\mathcal{C}(\tau ,2,%
\bar{n})$ is strongly antidegradable (weakly degradable) if and only if $%
\tau \leq 1/2$ ($\tau \geq 1/2$) \cite{VCH}.

\section{Quantum cryptography scenario}

In the standard scenario of quantum cryptography, the environment is
completely under control of a malicious eavesdropper (Eve). Here, a one-mode
channel can be generally seen as the effect of a collective attack, where
Eve probes the signals using individual interactions and then performs a
coherent detection of all the outputs collected in all the uses of the
channel. According to Ref.~\cite{GaussianATT}, one can define as a
\textquotedblleft canonical attack\textquotedblright\ a collective attack
that generates a one-mode Gaussian channel in canonical form. This is
actually a particular form of the most general collective Gaussian attack
that is completely characterized in Ref.~\cite{GaussianATT}. Up to partial
isometries, a canonical attack is described by combining the two-mode
Stinespring dilation $\{\mathbf{M}_{ae\tilde{e}},\left\vert 0\right\rangle \}
$ of the canonical form with the optimal coherent detection of all the
environmental outputs, which are collected in all the uses of the channel
[see Fig.~\ref{CharlieEvepic}(i) including Eve]. In the special case of the
class $B_{2}$, the corresponding $B_{2}$ canonical attacks are OGC attacks
where both the clone and anticlone are used in the final coherent
measurement. In all the other cases, the canonical attacks can be simplified
according to Fig.~\ref{CharlieEvepic}(ii) where Eve uses a single-mode
symplectic interaction $\mathbf{M}_{ae}$ and the TMSV state specified by
Eq.~(\ref{Input_EPR}). In particular, the \textit{regular canonical attacks}
are the ones with $\mathbf{M}_{ae}(\tau )$ given in Eq.~(\ref%
{Symplectic_Interaction}). These attacks can be associated to a pair $\{\tau
,\omega \}$ with $\tau \neq 0,1$.

In this paper, we consider the individual version of the regular canonical
attacks (denoted by $\{\tau ,\omega \}_{ind}$), where Eve is restricted to
incoherent detections of her outputs (and no isometry is applied). By
adopting this kind of attack, we derive the security thresholds of the
coherent state protocol of Ref.~\cite{Homo2}. In this protocol, Alice
prepares a coherent state $\rho _{a}:=\left\vert \alpha \right\rangle
\left\langle \alpha \right\vert $\ whose amplitude $\alpha $ is Gaussianly
modulated with variance $\mu $. Then, Alice sends the state through the
channel, whose output is homodyned by Bob. In particular, Bob randomly
switches between the detection of $\hat{q}_{b}$ and $\hat{p}_{b}$, the
effective sequence being classically communicated at the end of the protocol
(basis revelation). Here, the optimal attack $\{\tau ,\omega \}_{ind}$ is a
direct generalization of the delayed-choice entangling cloner attack of Ref.~%
\cite{Homo2,Estimators} (retrieved in the particular case $0<\tau <1$). This
means that Eve stores all her outputs in a quantum memory, awaits the basis
revelation and, then, performs the correct sequence of $\hat{q}$ and $\hat{p}
$ detections on her outputs. This is equivalent to saying that, for each run
of the protocol where Bob chooses the $\hat{q}$ quadrature, Eve also detects
the $\hat{q}$ quadrature on her modes $\{c,\tilde{c}\}$. In particular, we
can assume as first detection one of $\hat{q}_{\tilde{c}}$, which is
equivalent to the remote preparation of a $\hat{q}$-squeezed state on the
input mode $\{e\}$ with variance $\left\langle \hat{q}_{e}^{2}\right\rangle
=\omega ^{-1}$ \cite{Estimators}. As a consequence, Eve is always able to
control the input environment $\{e\}$ in such a way as to enhance her
detection of the output mode $\{c\}$ in the same quadrature which is
effectively chosen by Bob.

By adopting this optimal attack, let us explicitly derive the security
thresholds of the coherent state protocol in the limit of high modulation $%
\mu \rightarrow +\infty $. From Eqs.~(\ref{Input_EPR}) and~(\ref%
{Symplectic_Interaction}), we derive the following variance and conditional
variance for Bob's output%
\begin{equation}
V_{B}(\mu )=\left\langle \hat{q}_{b}^{2}\right\rangle =\left\langle \hat{p}%
_{b}^{2}\right\rangle =\left\vert \tau \right\vert (\mu +1)+\left\vert
1-\tau \right\vert \omega ~,~V_{B|A}=V_{B}(\mu =0)~.
\end{equation}%
Then, we have the following signal-to-noise formula for the classical mutual
information%
\begin{equation}
I_{AB}:=\frac{1}{2}\log \frac{V_{B}}{V_{B|A}}\overset{\mu \gg 1}{\rightarrow
}\frac{1}{2}\log \frac{\mu }{\eta (\omega ,\tau )}~,  \label{I_AB}
\end{equation}%
where the total noise $\eta (\omega ,\tau ):=\Delta +\chi (\omega ,\tau )$
is given by the sum of the quantum shot-noise $\Delta =1$ and the \emph{%
equivalent noise} of the channel%
\begin{equation}
\chi (\omega ,\tau )=\left\vert \frac{1-\tau }{\tau }\right\vert \omega ~.
\label{Equivalent_Noise}
\end{equation}%
From the point of view of the classical mutual information of Eq.~(\ref{I_AB}%
), the regular canonical form $\mathcal{C}(\tau ,2,\bar{n})$ is equivalent
to a form $\mathcal{C}(1,2,\bar{n})$ of the class $B_{2}$ (additive-noise
channel), where the input classical signal $\alpha $ with Gaussian
modulation $\mu $ is subject to the additive Gaussian noises $\chi $ and $%
\Delta $ (coming from the channel and the measurement, respectively). In
fact, in such a case, we would have%
\begin{equation}
V_{B}=\mu +\Delta +\chi ~,~V_{B|A}=\Delta +\chi ~,  \label{Add_descrip}
\end{equation}%
which leads exactly to Eq.~(\ref{I_AB}) for $\mu \rightarrow +\infty $. In
order to analyze the security thresholds, it is useful to introduce the
so-called \emph{excess noise}%
\begin{equation}
\varepsilon :=\eta (\omega ,\tau )-\eta (1,\tau )=\left\vert \frac{1-\tau }{%
\tau }\right\vert (\omega -1)
\end{equation}%
so that
\begin{equation}
\chi (\varepsilon ,\tau )=\left\vert \frac{1-\tau }{\tau }\right\vert
+\varepsilon ~,  \label{Chi_decomposition}
\end{equation}%
i.e., the equivalent noise can be decomposed in pure-$\tau $ noise and
excess noise $\varepsilon \geq 0$. Roughly speaking, $\varepsilon $
quantifies the effect of the input thermal noise ($\omega $) in the
equivalent additive description of the quantum channel, which is specified
by Eq.~(\ref{Add_descrip}).

\begin{figure}[tbph]
\vspace{-0.0cm}
\par
\begin{center}
\includegraphics[width=0.4\textwidth] {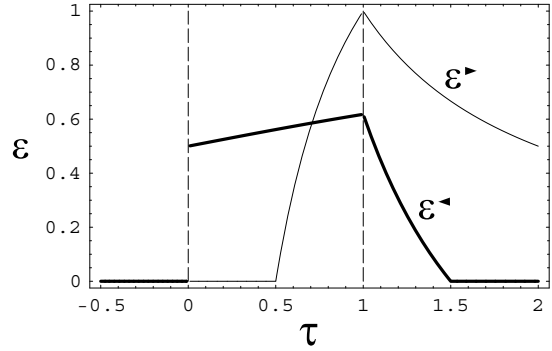}
\end{center}
\par
\vspace{-0.5cm}
\caption{Security thresholds in DR (thin curve) and RR (thick curve) in the
presence of an individual and regular canonical attack $\{\protect\tau ,%
\protect\omega \}_{ind}$ (where $\protect\tau \neq 0,1$). Such thresholds
are expressed in terms of maximum-tolerable excess noise $\protect%
\varepsilon $ versus $\protect\tau $. For a given $\protect\tau $, only the
positive $\protect\varepsilon ^{\prime }s$ below the curves are secure. }
\label{Sogliepic}
\end{figure}

In order to derive the security thresholds, let us compute the mutual
information $I_{AE}$ (between Alice and Eve) and $I_{BE}$ (between Bob and
Eve). It is easy to check that%
\begin{gather}
V_{E}(\mu )=\left\langle \hat{q}_{c}^{2}\right\rangle =\left\langle \hat{p}%
_{c}^{2}\right\rangle =\left\vert 1-\tau \right\vert (\mu +1)+\left\vert
\tau \right\vert \omega ^{-1}~, \\
V_{E|A}=\left\vert 1-\tau \right\vert +\left\vert \tau \right\vert \omega
^{-1}~,~V_{B|E}=\left[ \left\vert \tau \right\vert (\mu +1)^{-1}+\left\vert
1-\tau \right\vert \omega \right] ^{-1}~,
\end{gather}%
and, therefore,%
\begin{equation}
I_{AE}:=\frac{1}{2}\log \frac{V_{E}}{V_{E|A}}\overset{\mu \gg 1}{\rightarrow
}\frac{1}{2}\log \frac{\mu }{1+\chi ^{-1}}~,~I_{BE}:=\frac{1}{2}\log \frac{%
V_{B}}{V_{B|E}}\overset{\mu \gg 1}{\rightarrow }\frac{1}{2}\log \tau
^{2}\chi \mu ~.  \label{InfoEVE}
\end{equation}%
Then, we can compute the secret-key rates in direct reconciliation (DR, $%
\blacktriangleright $) and reverse reconciliation (RR, $\blacktriangleleft $%
), i.e.,%
\begin{equation}
R^{\blacktriangleright }:=I_{AB}-I_{AE}\rightarrow \frac{1}{2}\log \frac{%
1+\chi ^{-1}}{1+\chi }~,~R^{\blacktriangleleft }:=I_{AB}-I_{BE}\rightarrow
\frac{1}{2}\log \frac{1}{\tau ^{2}\chi (1+\chi )}~.
\end{equation}%
From $R^{\blacktriangleright }=0$ we derive the security threshold $\chi
(\varepsilon ,\tau )=1$ or, equivalently, the curve $\varepsilon
^{\blacktriangleright }=\varepsilon ^{\blacktriangleright }(\tau )$ shown in
Fig.~\ref{Sogliepic}. From such a figure we clearly see how strong
antidegradability (holding for $\tau \leq 1/2$) is a sufficient condition
for the insecurity of the channel in DR (since $\varepsilon
^{\blacktriangleright }=0$ for every $\tau \leq 1/2$). However, it is not a
necessary condition as shown by the existence of the insecure regions for $%
\tau \geq 1/2$ and $\varepsilon >\varepsilon ^{\blacktriangleright }(\tau )$
(where the channel is insecure but weakly degradable). This is a consequence
of the fact that Eve is much more powerful than Charlie, thanks to her
active control of the input environment. In fact, even if no strong
antidegradability can be found in the range $\tau \geq 1/2$, the channel can
be still antidegradable, e.g., within the insecure regions for $\tau \geq
1/2 $ and $\varepsilon >\varepsilon ^{\blacktriangleright }(\tau )$. We
recover a full equivalence between strong antidegradability and insecurity
only in the case $\varepsilon =0$, where the channel does not introduce
thermal noise. In such a case, in fact, the strong antidegradability
coincides with the standard antidegradability and the security threshold\ ($%
\tau =1/2$)\ corresponds exactly to the threshold between antidegradability
and degradability.

The fact that the strong antidegradability is a sufficient condition for the
insecurity in DR is quite obvious. In fact, it implies the
antidegradability, where Eve can reconstruct Bob's state and, therefore,
retrieve at least the same information of Bob in decoding Alice's signals
(i.e., $\exists \mathcal{A}^{s}\Rightarrow \exists \mathcal{A}\Rightarrow
I_{AE}\geq I_{AB}$). However, the situation is completely different in RR,
where Alice and Eve try to guess Bob's outcomes. In such a case, even if the
channel is strongly antidegradable, Bob's outcomes can be much more
correlated to Alice's variables than Eve's ones. In general, the only way
for Eve to beat Alice in RR\ consists in introducing an environment which is
squeezed enough to make her correlations prevail. From $R^{%
\blacktriangleleft }=0$ we derive the discontinuous \cite{Discontinuity}
security threshold
\begin{equation}
\varepsilon ^{\blacktriangleleft }=\varepsilon ^{\blacktriangleleft }(\tau
):=\frac{\sqrt{4+\tau ^{2}}-\left\vert \tau \right\vert -2\left\vert 1-\tau
\right\vert }{2\left\vert \tau \right\vert }~,
\end{equation}%
shown in Fig.~\ref{Sogliepic}. From Fig.~\ref{Sogliepic} it is clear that,
even if the channel is strongly antidegradable, QKD can be secure. This is
due to the existence of the secure region for $0<\tau \leq 1/2$ and $%
\varepsilon <\varepsilon ^{\blacktriangleleft }(\tau )$. Notice that for $%
\tau >1$, i.e., for an amplifying channel, reverse reconcilation is
outperfomed by direct reconciliation. This is in accordance with the
previous results of Ref.~\cite{Radim}.

According to the expression of $I_{AE}$ in Eq.~(\ref{InfoEVE}), the
Alice-Eve channel can also be described by an additive-noise channel where
the input classical signal $\alpha $\ (with variance $\mu $) is modulated by
an equivalent channel's noise $\chi ^{-1}$ and a homodyne detection noise $%
\Delta =1$. In fact, we retrieve the same mutual information $I_{AE}$ of
Eq.~(\ref{InfoEVE}) by considering
\begin{equation}
V_{E}=\mu +\Delta +\chi ^{-1}~,~V_{E|A}=\Delta +\chi ^{-1}~,
\end{equation}%
and taking the asymptotic limit for $\mu \rightarrow +\infty $. By
considering both the Alice-Bob and Alice-Eve channels, one easily checks
that the optimal $\{\tau ,\omega \}_{ind}$ has therefore an equivalent
additive description when direct reconciliation and high modulation are
considered. Such an additive description corresponds to an individual OGC
attack where Eve clones the input signals with cloning variances $\chi
_{b}=\chi $ and $\chi _{c}=\chi ^{-1}$, stores her clones in a quantum
memory and, then, makes the correct homodyne detections after the basis
revelation \cite{Anticlone}. Such an individual attack is optimal since the
saturation of the uncertainty principle $\chi _{b}\chi _{c}=1$ minimizes the
information-disturbance trade-off, which can be expressed by the product of
the output conditional variances
\begin{equation}
V_{B|A}V_{E|A}=(\Delta +\chi )(\Delta +\chi ^{-1})~.
\end{equation}%
In direct reconciliation and high modulation, an individual OGC attack with
noise $\chi $ represents therefore an equivalent additive description of the
optimal attack $\{\tau ,\omega \}_{ind}$ via Eq.~(\ref{Equivalent_Noise}).
To be precise, there is a whole class of optimal attacks $\{\tau ,\omega
\}_{ind}$, with different $\tau $ and $\omega $ but the same $\chi =\chi
(\omega ,\tau )$, which are equivalent to an individual OGC\ attack. Notice
that this equivalence is true for the \textquotedblleft
switching\textquotedblright\ protocol of Ref.~\cite{Homo2}, but not for the
\textquotedblleft non-switching\textquotedblright\ protocol of Ref.~\cite%
{Hetero2}.

\section{Conclusion}

In this paper, we have investigated recent notions and properties of the
one-mode Gaussian channels in the scenario of quantum cryptography. In
particular, we have considered the canonical attacks, which are the
cryptographic analog of the canonical forms. We have adopted the individual
version of these attacks in order to study the connections between the
degradability properties of a Gaussian channel and its security for QKD. We
have also explicitly clarified the connections between the various notions
of degradability and antidegradability. Finally, we have shown some
connections between individual canonical attacks and optimal Gaussian
cloners.

\section{Acknowledgments}

S.P. was supported by a Marie Curie Fellowship of the European Community.
S.L. was supported by the W.M. Keck foundation center for extreme quantum
information theory (xQIT).


\begin{thebibliography}{99}
\bibitem{PRnote} A physical representation is a unitary dilation of the
quantum channel where the environmental state $\rho _{E}$ can be (generally)
mixed \cite{VCH} (see also Ref.~\cite{PR}). It is \textit{not unique} up to
partial isometries, except when it coincides with a Stinespring dilation
(i.e., $\rho _{E}$ is pure).
\bibitem{B2} The class $B_{2}$ is the unique class which is neither
anti-degradable nor degradable \cite{HOL}.
\bibitem{Anticlone} Notice that, in order to be optimal in DR, the
individual OGC\ attack ignores the measurement of the anticlone. It is not
known if such a further measurement can be useful in the eavesdropping of
the protocol in RR.
\end{thebibliography}
\end{document}